# $N = 2$ Supersymmetric Calabi-Yau Hypersurface Sigma-Models on Curved Two-Dimensions[1]

Hitoshi NISHINO[2]

*Department of Physics and Astronomy*
*Howard University*
*Washington, D.C. 20059, USA*

and

*Department of Physics*
*University of Maryland at College Park*
*College Park, MD 20742-4111, USA*

**Abstract**

We consider the effect of curved two-dimensional space-time on Witten's $N = 2$ supersymmetric sigma models interpolating Calabi-Yau hypersurfaces to Landau-Ginzburg models. In order for the former models to have significant connection to superstring theory, only the $N = (1, 1)$ or $N = (1, 0)$ part of the total $N = (2, 2)$ world-sheet supersymmetry is made local. Even though there arises an additional minimizing condition due to a scalar auxiliary field in the supergravity multiplet on curved two-dimensions, the essential feature of the sigma-model relating Calabi-Yau and Landau-Ginzburg models will be maintained. This indicates the validity of these sigma models formulated on curved two-dimensions or curved world-sheets. As a by-product, the coupling of $N = (2, 2)$ vector multiplets to other multiplets with $N = (1, 1)$ local supersymmetry is developed.

---

[1]This work is supported in part by DOE grant # DE-FG02-94ER40854
[2]E-Mail: Nishino@umdhep.umd.edu

## 1. Introduction

In 1985, what is called Calabi-Yau (CY) manifold was first recognized as one of the most important spaces for phenomenological model building for superstring theories [1][2]. It seems impossible to avoid these CY spaces, as long as the target supersymmetry is maintained intact up to some energy level. In terms of world-sheet physics, a superstring model with such a CY target space should be equivalent to a global $N=2$ sigma-model with *local* $N=1$ supersymmetry with the target space CY metric [2][3]. Eventually many physically interesting properties of phenomenological models should be computable based on this CY geometrical $\sigma$-model within two-dimensions $(D=2)$.

However, there is a practical drawback of this CY geometrical sigma-model approach [3] due to the lack of manifest conformal invariance with non-linear structures at the quantum level. For example, the renormalization flow analysis in these models are difficult to handle beyond the one-loop level, and definitely we need some simpler formulation. There was also need for a nice formulation that can provide a clear link between the CY geometrical sigma-models and Landau-Ginzburg (LG) theory [4], because these theories share many similarities. In a recent approach by Witten [5], these needs have been desirably satisfied by considering an $N=2$ supersymmetric sigma-model with constraints acting on linear field equations, which result in a hypersurface that is equivalent to a CY manifold. As desired, this sigma-model describes not only the CY manifold but also the LG theory, as two different parameter ranges of the same $D=2$ theory as analytic continuation. Furthermore, such a linearization approach [5] makes it easy to deal with constrained CY hypersurfaces in Grassmannians [6] and gauged LG orbifolds corresponding to toric varieties, or more complicated models.

This type of sigma-models so far always dealt with $N=2$ *global* supersymmetry on the world-sheet. However, it is imperative to consider the *curved* world-sheet in order for these models to be appropriate superstring theory, and therefore we have to consider *local* supersymmetry in $D=2$. Moreover, the models in ref. [5] would correspond to the CY geometrical models [3] as the low-energy limit, upon the inclusion of local supersymmetry. To be consistent with the target space-time supersymmetry, we have to make only the $N=(1,1)$ or $N=(1,0)$ supersymmetry *local* out of the initial $N=2$ supersymmetries, while keeping the total $N=2$ supersymmetry *global* aside from the supergravity-couplings. The case of $N=(1,1)$ local supersymmetry corresponds to the superstring with $N=2$ space-time supersymmetry [7], while $N=(1,0)$ corresponds to $N=1$ space-time supersymmetry [8].

An interesting question then is whether or not those particular constraints specifying the CY hypersurfaces are affected by the curved world-sheet, especially by couplings to the $N=(1,1)$ or $N=(1,0)$ *local* supersymmetry. In this paper we try to answer this question by explicit couplings of Poincaré[3] supergravity to those global $N = (2, 2)$ models. We will see that the main properties of the global models are intact even with the couplings to the $N = (1, 1)$ or $N = (1, 0)$ supergravity, and therefore the validity and appropriateness of these models as phenomenological superstring models on general curved world-sheets will be maintained as the global case.

## 2. Invariant Lagrangians

For our purpose, we need to establish the the appropriate $N = (1,1)$ local supersymmetry couplings [9] to arbitrary number of chiral multiplets and $U(1)$ vector multiplets [10], keeping the global $N = (2,2)$ supersymmetry before the supergravity couplings. We basically use notations similar to Witten's paper [5] for these multiplets.

For the first illustration of our chiral multiplet consisting of the component fields $(\phi^A, \psi^A, F^A)$, (A, B, ⋯ = 1, 2), we give their Poincaré supertranslation rules:

$$
\begin{aligned}
\delta\phi^A &= \bar{\epsilon}^1\psi^A - \epsilon^{AB}\bar{\epsilon}^2\psi^B \ , \\
\delta\psi^A &= -i\gamma^\mu\epsilon^1\widehat{D}_\mu\phi^A - i\epsilon^{AB}\gamma^\mu\epsilon^2\widehat{D}_\mu\phi^B + \epsilon^1 F^A - \epsilon^{AB}\epsilon^2 F^B \\
&\quad + Q\left(\epsilon^1\phi^A\sigma^2 + \epsilon^{AB}\epsilon^2\phi^B\sigma^2 - \gamma_5\epsilon^2\phi^A\sigma^1 + \epsilon^{AB}\gamma_5\epsilon^1\phi^B\sigma^1\right) - \nu_{\rm C}\epsilon^1 S\phi^A \ , \\
\delta F^A &= -i\bar{\epsilon}^1\gamma^\mu\widehat{D}_\mu\psi^A - i\epsilon^{AB}\bar{\epsilon}^2\gamma^\mu\widehat{D}_\mu\psi^B \\
&\quad + Q\Big(\epsilon^{AB}\bar{\epsilon}^1\lambda^1\phi^B - \bar{\epsilon}^1\lambda^2\phi^A + \epsilon^{AB}\bar{\epsilon}^1\gamma_5\psi^B\sigma^1 - \bar{\epsilon}^1\psi^A\sigma^2 \\
&\quad\quad - \bar{\epsilon}^2\lambda^1\phi^A - \epsilon^{AB}\bar{\epsilon}^2\lambda^2\phi^B - \bar{\epsilon}^2\gamma_5\psi^A\sigma^1 - \epsilon^{AB}\bar{\epsilon}^2\psi^B\sigma^2\Big) + \nu_{\rm C}\delta(S\phi^A) \ . \quad (2.1)
\end{aligned}
$$

We chose each of the fermionic fields $\psi^A$ to be Majorana, while all the bosonic fields $\phi^A$, $F^A$ to be real. The scalar field $S$ is an auxiliary field for the $N = (1,1)$ supergravity multiplet to be given later. We use the Latin indices $m, n, \cdots = (0), (1)$ for the local Lorentz indices, while the Greek ones $\mu, \nu, \cdots = 0, 1$ for the general coordinates. Our signature is $\eta_{mn} = {\rm diag.}(+,-)$, and $\{\gamma_m, \gamma_n\} = 2\eta_{mn}$. Relevantly, we have $\gamma^{\mu\nu} = e^{-1}\epsilon^{\mu\nu}\gamma_5$ for $\gamma_5 \equiv \gamma^{(0)}\gamma^{(1)}$. The constant $Q$ is the $U(1)$-coupling constant, and the $\epsilon^{AB}$-tensor is the usual one: $\epsilon^{12} = -\epsilon^{21} = +1$, and the *bars* on fermions are similar to those for Majorana fermions for $D = 4, N = 1$ supersymmetry [11]. At later stages, we will also use the complex notation by the identification

$$
\begin{aligned}
\phi &\equiv \tfrac{1}{\sqrt{2}}\left(\phi^1 + i\phi^2\right) \ , & \phi^* &\equiv \tfrac{1}{\sqrt{2}}\left(\phi^1 - i\phi^2\right) \ , \\
F &\equiv \tfrac{1}{\sqrt{2}}\left(F^1 + iF^2\right) \ , & F^* &\equiv \tfrac{1}{\sqrt{2}}\left(F^1 - iF^2\right) \ , \quad (2.2) \\
\psi &\equiv \tfrac{1}{\sqrt{2}}\left(\psi^1 + i\psi^2\right) \ , & \overline{\psi} &\equiv \tfrac{1}{\sqrt{2}}\left(\psi^1 - i\psi^2\right) \ .
\end{aligned}
$$

---
[3]The original hypersurface models themselves [5] are not superconformally invariant in general, *except for* infrared fixed points. Hence we consider only Poincaré supergravity.

In the last expression, both $\psi$ and $\overline{\psi}$ are Weyl spinors complex conjugate to each other, and the meaning of the *bar* is different from an inner product for two Majorana spinors. Since we are making only the $N = (1,1)$-part of the total $N = (2,2)$ local in this section, the conformal or Poincaré supergravity multiplets we need are those for $N = (1,1)$ local supersymmetry. Relevantly, we see that the global $N = (2,2)$ symmetry is violated by the supercovariant derivatives only for the first supersymmetry out of $N = 2$. Accordingly, the Poincaré supercovariant derivative $\widehat{D}_\mu$ is covariant under the $U(1)$ gauge, Lorentz, and $N = (1,1)$ Poincaré local supersymmetry in $D = 2$. For example,

$$\widehat{D}_\mu \phi_i^A \equiv \partial_\mu \phi_i^A + Q_{i,a} \epsilon^{AB} V_{\mu a} \phi_i^B - \overline{\psi}_\mu \psi_i^A \quad , \tag{2.3}$$

where $V_{\mu a}$ is the $U(1)$ gauge field, $\psi_\mu$ is the $N = (1,1)$ gravitino, and $\nu_{\rm C}$ is the conformal weight of the chiral multiplet. Relevantly, the *hat*-symbols are commonly used to any on quantities that are Poincaré supercovariant. The indices $i, j, \cdots$ and $a, b, \cdots$ distinguish different chiral multiplets and distinct $U(1)$ vector multiplets respectively with mutual minimal coupling constants $Q_{i,a}$ as in ref. [5]. The local supercovariance is only for the parameter $\epsilon^1$ out of $\epsilon^A$. For convenience sake as well as to accord with the standard canonical multiplet with manifest conformal invariance, we choose

$$\nu_{\rm C} = 0 \quad . \tag{2.4}$$

All the $Q$-dependent terms in (2.1) can be easily fixed by combining the results for $N = (1,1)$ local supersymmetry [9][12] with the global case [5]. The structure of these terms is also essentially similar to the $D = 4$ case tensor calculus [11]. As is easily seen, if we switch-off the effect of supergravity, (2.1) will be reduced to the chiral multiplet with the global $N = (2,2)$ supersymmetry [5].

The multiplet of $N = (1,1)$ supergravity $(e_\mu{}^m, \psi_\mu, S)$ transforms under Poincaré supersymmetry as [9][12]

$$\begin{aligned} \delta e_\mu{}^m &= -2i\,\overline{\epsilon}^1 \gamma^m \psi_\mu \quad , \\ \delta \psi_\mu &= D_\mu \epsilon^1 - \frac{i}{2} \gamma_\mu \epsilon^1 S \quad , \\ \delta S &= 2\overline{\epsilon}^1 \gamma^{\mu\nu} D_\mu \psi_\nu + i \overline{\epsilon}^1 \gamma^\mu \psi_\mu S \equiv 2\overline{\epsilon}^1 \gamma^{\mu\nu} \widehat{D}_\mu \psi_\nu \quad . \end{aligned} \tag{2.5}$$

Note that we have only one auxiliary field $S$ in this multiplet due to the $N = (1,1)$ local supersymmetry. When we deal with the $N = (1,0)$ supergravity later, we simply impose the positive handed-ness on the parameter $\epsilon^1$, and accordingly the auxiliary field $S$ will disappear, forming the irreducible $N = (1,0)$ supergravity multiplet $(e_\mu{}^m, \psi_\mu^+)$ [13][14].

We can now give the super Poincaré transformation rules for the $U(1)$ vector multiplet

$(V_\mu, \lambda^A, \sigma^A, D)^4$

$$\delta V_\mu = i\bar{\epsilon}^A \gamma_\mu \lambda^A - 2\bar{\epsilon}^1 \gamma_5 \psi_\mu \sigma^1 \ ,$$

$$\delta\lambda^1 = \tfrac{1}{2}\gamma^{\mu\nu}\epsilon^1 \widehat{V}_{\mu\nu} - i\gamma_5\gamma^\mu \epsilon^1 \widehat{D}_\mu \sigma^1 + i\gamma^\mu \epsilon^2 \widehat{D}_\mu \sigma^2 + \epsilon^2 D - \nu_1 \gamma_5 \epsilon^1 S\sigma^1 \ ,$$

$$\delta\lambda^2 = \tfrac{1}{2}\gamma^{\mu\nu}\epsilon^2 \widehat{V}_{\mu\nu} - i\gamma^\mu \epsilon^1 \widehat{D}_\mu \sigma^2 - i\gamma_5\gamma^\mu \epsilon^2 \widehat{D}_\mu \sigma^1 - \epsilon^1 D - \nu_1 \epsilon^1 S\sigma^2 \ , \qquad (2.6)$$

$$\delta\sigma^1 = \bar{\epsilon}^A \gamma_5 \lambda^A \ , \qquad \delta\sigma^2 = \epsilon^{AB}\bar{\epsilon}^A \lambda^B \ ,$$

$$\delta D = i\epsilon^{AB}\bar{\epsilon}^A \widehat{\slashed{D}}\lambda^B - \nu_1 \delta(\sigma^2 S) \ .$$

When there are plural vector multiplets, we will distinguish them by the subscripts $a$, $b$, $\cdots$. We will also use the complex field $\sigma \equiv (\sigma^1 + i\sigma^2)/\sqrt{2}$ later.

There are a few remarks in order. To get this $N = (2,2)$ vector multiplet, we have combined one $N = (1,1)$ vector multiplet [10] and one $N = (1,1)$ chiral multiplet whose supercovariances are established [9][12]. It turned out that when the conformal weight is

$$\nu_1 = -1 \ , \qquad (2.7)$$

we can conveniently identify the $F$-component of the $N = (1,1)$ chiral multiplet with $(1/2)\epsilon^{-1}\epsilon^{\mu\nu}\widehat{V}_{\mu\nu}$, and construct an invariant lagrangian with the $U(1)$-couplings. As a matter of fact, this is shown to be the only possible choice in terms of superspace language [10].

We can now proceed to the invariant lagrangians for these multiplets, as in the global case [5]. They are respectively that of chiral multiplets, that of general superpotential, that of $D$-type and topological term, that of vector multiplets, and that of twisted superpotential [5][15], respectively abbreviated as CM, $W$, $D$, $\theta$, VM and $\widetilde{W}$. Their explicit forms are

$$e^{-1}\mathcal{L}_{\text{CM}} = \sum_i \left[ + \tfrac{1}{2}(D_\mu \phi_i^A)^2 + \tfrac{i}{2}\overline{\psi}_i^A \gamma^\mu D_\mu \psi_i^A + \tfrac{1}{2}(F_i^A)^2 - \overline{\psi}_\mu \gamma^\nu \gamma^\mu \psi_i^A \left(D_\nu \phi_i^A - \tfrac{1}{2}\overline{\psi}_\nu \psi_i^A\right) \right.$$
$$+ \sum_a \left\{ Q_{i,a}(\overline{\lambda}_a^2 \psi_i^A)\phi_i^A - Q_{i,a}\epsilon^{AB}(\overline{\lambda}_a^1 \psi_i^A)\phi_i^B + \tfrac{1}{2}Q_{i,a}\sigma_a^2(\overline{\psi}_i^A \psi_i^A) - Q_{i,a}\sigma_a^1(\overline{\psi}_i^1 \gamma_5 \psi_i^2) \right.$$
$$+ Q_{i,a}D_a|\phi_i|^2 - 2Q_{i,a}^2|\sigma_a|^2|\phi_i|^2 + iQ_{i,a}\overline{\psi}_\mu \gamma^\mu \lambda_a^2|\phi_i|^2 + Q_{i,a}\overline{\psi}_\mu \gamma^{\mu\nu}\psi_\nu |\phi_i|^2 \sigma_a^2$$
$$\left. \left. + iQ_{i,a}\overline{\psi}_\mu \gamma^\mu \psi_i^A \phi_i^A \sigma_a^2 - iQ_{i,a}\epsilon^{AB}\overline{\psi}_\mu \gamma_5 \gamma^\mu \psi_i^A \phi_i^B \sigma_a^1 \right\} \right] \ , \qquad (2.8)$$

$$e^{-1}\mathcal{L}_W = \left[ \sum_i (-F_i + i\overline{\psi}_\mu \gamma^\mu \psi_i) \frac{\partial W}{\partial \phi_i} + \tfrac{1}{2}\sum_{i,j}(\psi_i \psi_j)\frac{\partial^2 W}{\partial \phi_i \partial \phi_j} - \overline{\psi}_\mu \gamma^{\mu\nu}\psi_\nu W - SW \right] + \text{h.c.} \ , \qquad (2.9)$$

---

[4]The global $N = (2,2)$ vector multiplet was first given ref. [10].

$$e^{-1}\mathcal{L}_{D,\theta} = -\sum_a r_a \left[ D_a + i\overline{\psi}_\mu \gamma^\mu \lambda_a^2 - \sigma_a^2 \overline{\psi}_\mu \gamma^{\mu\nu} \psi_\nu - 2S\sigma_a^2 \right] + \sum_a \frac{\theta_a}{4\pi} e^{-1} \epsilon^{\mu\nu} V_{\mu\nu\,a} \quad , \quad (2.10)$$

$$\begin{aligned} e^{-1}\mathcal{L}_{\text{VM}} = \sum_a \frac{1}{\widetilde{e}_a^2} \Big[ & -\tfrac{1}{4}(\widehat{V}_{\mu\nu\,a})^2 + \tfrac{i}{2}\overline{\lambda}_a^A \gamma^\mu D_\mu \lambda_a^A + \tfrac{1}{2}(\partial_\mu \sigma_a^A)^2 + \tfrac{1}{2} D_a^2 \\ & - \overline{\psi}_\mu \gamma_5 \gamma^\nu \gamma^\mu \lambda_a^1 \left( \partial_\nu \sigma_a^1 - \tfrac{1}{2} \overline{\psi}_\nu \gamma_5 \lambda^1 \right) - \overline{\psi}_\mu \gamma^\nu \gamma^\mu \lambda_a^2 \left( \partial_\nu \sigma_a^2 - \tfrac{1}{2} \overline{\psi}_\nu \lambda_a^2 \right) \\ & + S^2 |\sigma_a|^2 + S \left( \tfrac{1}{2} e^{-1} \epsilon^{\mu\nu} \widehat{V}_{\mu\nu\,a} \sigma_a^1 - \sigma_a^2 D_a \right) \Big] \quad , \quad (2.11) \end{aligned}$$

$$\begin{aligned} e^{-1}\mathcal{L}_{\widetilde{W}} = \Bigg[ & \sqrt{2} \sum_a \left( D_a - \tfrac{i}{2} e^{-1} \epsilon^{\mu\nu} \widehat{V}_{\mu\nu\,a} \right) \frac{\partial \widetilde{W}}{\partial \widetilde{\sigma}_a} - 2S \left( \sum_a \widetilde{\sigma}_a \frac{\partial \widetilde{W}}{\partial \widetilde{\sigma}_a} + \widetilde{W} \right) \\ & + \sum_{a,b} (\widetilde{\lambda}_a \widetilde{\lambda}_b) \frac{\partial^2 \widetilde{W}}{\partial \widetilde{\sigma}_a \partial \widetilde{\sigma}_b} + 2i \sum_a \overline{\psi}_\mu \gamma^\mu \widetilde{\lambda}_a \frac{\partial \widetilde{W}}{\partial \widetilde{\sigma}_a} - 2 \overline{\psi}_\mu \gamma^{\mu\nu} \psi_\nu \widetilde{W} \Bigg] + \text{h.c.} \quad (2.12) \end{aligned}$$

In (2.9) the symbol $(\psi\psi)$ signifies the inner product of two Weyl spinors $\psi$ defined by (2.2), which is equivalent to $(1/2)\overline{\psi}^A \psi^A$. From now on, whenever $\psi$ has no index $A, B, \cdots$, it is supposed to be a Weyl spinor, e.g., in the term $\overline{\psi}_\mu \gamma^\mu \psi$ in (2.9), $\psi$ denotes a Weyl spinor. The indices $i, j, \cdots$ and $a, b, \cdots$ are respectively for distinct chiral and vector multiplets. As usual, the superpotential $W$ is holomorphic: $\partial W / \partial \phi^* = 0$. In (2.10) we allow each $U(1)$ vector multiplet to have different couplings $r_a$ and $\theta_a$ with the suffix $a$. In (2.12), the *tilded* fields are defined by

$$\widetilde{\sigma} \equiv \frac{1}{\sqrt{2}} \left( \sigma^2 + i\sigma^1 \right) \quad , \quad \widetilde{\lambda} \equiv \frac{1}{\sqrt{2}} \left( \lambda^2 + i\gamma_5 \lambda^1 \right) \quad . \quad (2.13)$$

The switched positions of $\sigma^1$ and $\sigma^2$ is for the appropriate parity, such that the imaginary part is to be pseudo-scalar. Accordingly, $\widetilde{W}$ is a holomorphic function only of $\widetilde{\sigma}_a$: $\partial \widetilde{W} / \partial \widetilde{\sigma}_a^* \equiv 0$. These *tildes* are also used as the reminder of the twisted superpotential $\widetilde{W}$ distinguished from $W$.

The validity of our lagrangians above can be easily reconfirmed, e.g., by deriving all the field equations in the system, and see their automatic supercovariantization. This is because if the total lagrangian is Poincaré superinvariant, all the field equation should be supercovariant after arranging all the relevant terms. Needless to say, all the global parts of the above lagrangians agree with the global results in ref. [5].

As the supertranslation rule (2.1) was similar to the $D = 4$ case, our lagrangians have resemblance to the corresponding $D = 4$ cases [11]. For example, there are $Q$-dependent couplings of the form $\psi_\mu \lambda^2 |\phi|^2$. The $\psi_\mu \psi \phi \sigma$-type terms correspond to the Noether couplings $\psi_\mu \psi^A D_\nu \phi$, because the $\sigma^A$-fields corresponds to the extra dimensional components of $A_\mu$ when the $D = 4$ expressions are reduced into $D = 2$. However, an important difference of (2.8) from its $D=4$ analog is the absence of a supergravity lagrangian multiplied by a scalar function of Brans-Dicke type [9]. The absence of the term with the gravitino strength $\gamma^{\mu\nu}\widehat{D}_\mu\psi_\nu$ also corresponds to this feature.

Even though some of these invariant lagrangians were given a long time ago, *e.g.*, the global $N=(2,2)$ vector multiplets [10], or local tensor calculus for $N=(1,1)$ chiral multiplets [12], or $N=(2,0)$ heterotic sigma model coupled to $N=(2,0)$ conformal supergravity [13], we emphasize that the $N=(1,1)$ supergravity coupled to $N=(2,2)$ vector multiplets with the $Q$-dependent minimal $U(1)$-gauging in the context of hypersurface sigma-models are the important new results here.

## 3. Bosonic Potential and Effect of $N=(1,1)$ Supergravity

Since we have already established the couplings of $N=(1,1)$ supergravity to the sigma-model, we can now look into the question of the effect of local supergravity on the bosonic potential terms.

To this end, we can concentrate only on the purely bosonic terms, ignoring all the terms with fermions. First of all, we notice the usual fact that the metric and the gravitino field equations out of the supergravity multiplet give the super Virasoro conditions [2]. At the classical level, this is the usual constraint equation for the energy-momentum tensor and spinor current to vanish [2], and moreover their importance is always related to the kinetic terms, which are separate from the potential terms we are going to deal with.[5] From this viewpoint, we do not worry about the consistency of the field equations for the zweibein and gravitino. Relevantly, we do not include the twisted lagrangian (2.12) in the analysis below.

After these considerations, we get the following lagrangian for the bosonic potential:

$$
\begin{aligned}
e^{-1}\mathcal{L}_{\text{Pot}} &= \sum_i \left[ |F_i|^2 + \sum_a \left( Q_{i,a} |\phi_i|^2 D_a - 2 Q_{i,a}^2 |\sigma_a|^2 |\phi_i|^2 \right) \right] \\
&\quad - \left[ \sum_i F_i \frac{\partial W}{\partial \phi_i} + SW + \text{h.c.} \right] - \sum_a r_a \left( D_a - 2S\sigma_a^2 \right) + \sum_a \frac{\theta_a}{4\pi} e^{-1} \epsilon^{\mu\nu} V_{\mu\nu\,a} \\
&\quad + \sum_a \frac{1}{\widetilde{e}_a^2} \left[ \tfrac{1}{2}(D_a)^2 + S^2|\sigma_a|^2 + \tfrac{1}{2} e^{-1}\epsilon^{\mu\nu} V_{\mu\nu\,a} S\sigma_a^1 - S\sigma_a^2 D_a \right] - \frac{1}{4}\sum_a \frac{1}{\widetilde{e}_a^2} (V_{\mu\nu\,a})^2 \quad .
\end{aligned} \tag{3.1}
$$

The terms with $V_{\mu\nu}$ are also included here due to their mixture with other fields. At this stage, we already see that the bosonic lagrangian (3.1) seems very similar to the global case [5], except for the only effect by supergravity *via* the extra terms with the auxiliary field $S$. We first eliminate the auxiliary fields $F$, $D$ and $S$ by forming perfect squares, as usual:

$$
\begin{aligned}
e^{-1}\mathcal{L}_{\text{Pot}} = &-\left[\sum_i\left|\frac{\partial W}{\partial \phi_i}\right|^2 + \tfrac{1}{2}\sum_a \widetilde{e}_a^2\left(\sum_i Q_{i,a}|\phi_i|^2 - r_a\right)^2 + 2\sum_{i,a} Q_{i,a}^2|\sigma_a|^2|\phi_i|^2 + \sum_a \frac{\widetilde{e}_a^2 \widetilde{\theta}_a^2}{8\pi^2}\right.\\
&\left. + \left\{\sum_b \frac{2(\sigma_b^1)^2}{\widetilde{e}_b^2}\right\}^{-1}\left\{W + W^* - \sum_a\left(\sum_i Q_{i,a}|\phi_i|^2\sigma_a^2 + r_a\sigma_a^2 + \frac{1}{2\widetilde{e}_a^2}e^{-1}\epsilon^{\mu\nu}V_{\mu\nu\,a}\sigma_a^1\right)\right\}^2\right]\\
&+ \sum_i\left|F_i - \frac{\partial W}{\partial \phi_i}\right|^2 + \tfrac{1}{2}\sum_a \frac{1}{\widetilde{e}_a^2}\left(D_a - S\sigma_a^2 + \widetilde{e}_a^2\sum_i Q_{i,a}|\phi_i|^2 - \widetilde{e}_a^2 r_a\right)^2\\
&+ \sum_c \frac{(\sigma_c^1)^2}{2\widetilde{e}_c^2}\left[S + \left\{\sum_b \frac{(\sigma_b^1)^2}{\widetilde{e}_b^2}\right\}^{-1}\left\{W + W^* - \sum_a\left(\sum_i Q_{i,a}|\phi_i|^2\sigma_a^2 + r_a\sigma_a^2 + \frac{1}{2\widetilde{e}_a^2}e^{-1}\epsilon^{\mu\nu}V_{\mu\nu\,a}\sigma_a^1\right)\right\}\right]^2\\
&+ \sum_a \frac{1}{8\widetilde{e}_a^2}\left(e^{-1}\epsilon^{\mu\nu}V_{\mu\nu\,a} - \frac{\widetilde{e}_a^2 \widetilde{\theta}_a}{\pi}\right)^2\ . \quad (3.2)
\end{aligned}
$$

---
[5]The only exception is the contribution from the potential to the trace of the energy-momentum tensor. We will come back to this later.

The $\widetilde{\theta}$ ($-\pi \leq \widetilde{\theta} < \pi$) is different from the original $\theta$ only by $2\pi n$ ($n \in \mathbb{Z}$) [5]. The first two lines in (3.2) are for the manifestly positive definite potential terms. The remaining three lines show the field equations in perfect square forms for the auxiliary fields $F$, $D$ and $S$, as well as for $V_{\mu\nu}$ reproducing the global case result $V^{\mu\nu} = -\widetilde{e}^2 \widetilde{\theta} e^{-1}\epsilon^{\mu\nu}/(2\pi)$ [5] with generally non-zero minimum, as the last term of the first line shows. In our locally supersymmetric case, since a non-zero minimum energy contributes to the energy-momentum tensor which is forced to vanish by the zweibein field equation, it seems that $\widetilde{\theta} = 0$ is the only acceptable solution when supergravity is switched on. The second line in (3.2) is a new term coming from the elimination of the auxiliary field $S$ which can be minimized to zero value, when

$$(W + W^*) - \sum_a\left(\sum_i Q_{i,a}\sigma_a^2|\phi_i|^2 + r_a\sigma_a^2 + \frac{1}{2\widetilde{e}^2}e_a^{-1}\epsilon^{\mu\nu}V_{\mu\nu\,a}\sigma_a^1\right) = 0\ . \quad (3.3)$$

The relative sign between the $Q\sigma^2|\phi|^2$-term and $r\sigma^2$-term in (3.3) is opposite to that in the second term in (3.2), as will be important later.

## 4. Couplings to $N = (1,0)$ Supergravity

Once we have completed the $N = (1,1)$ supergravity couplings, it is easy to deal with the $N = (1,0)$ supergravity couplings [8][13][14]. This is because the $N = (1,0)$ supergravity multiplet $(e_\mu{}^m, \psi_\mu^+)$ is a sub-multiplet of the $N = (1,1)$ supergravity multiplet, and we can simply truncate the left-handed part of the gravitino and the auxiliary field in the latter to get the former. Accordingly, some terms in the transformation rules in (2.1) - (2.6) disappear, as well as terms in the lagrangians (2.8) - (2.12), whose explicit forms are skipped here, because one can easily reproduce them. All the other field contents such as those of the chiral multiplet and the vector multiplet will be maintained. The most crucial point in this truncation is that all the terms with the auxiliary field $S$ disappear completely, and eventually we have no extra condition (3.3) in the $N = (1,0)$ supergravity couplings. The absence of the extra condition for $N = (1,0)$ supergravity couplings reflects the fact that $D = 10$, $N = 1$ superstring theory is *less* restrictive than $D = 10$, $N = 2$ superstring.[6]

## 5. CY–LG Correspondence on Curved World-Sheet

We can now analyze the effect by curved world-sheet on an explicit sigma-models [5] interpolating the CY hypersurfaces and LG models. Following ref. [5], we introduce $n$-copies of chiral superfields $S_i$ $(i, j, \cdots = 1, \cdots, n)$ and an extra chiral superfield $P$, together with an $U(1)$ vector multiplet with the coupling constant $\widetilde{e}$. In order to avoid the $U(1)$ anomaly [5][16], we have to satisfy the relation

$$\sum_i Q_{i,a} = 0 \quad (\forall a) \; . \tag{5.1}$$

Hence we assign the $U(1)$ charges of $S_i$ to be $Q_i = +1$, while that of $P$ to be $Q_0 = -n$ [5]. The superpotential is[7]

$$W = P \cdot G(S_i) \; , \tag{5.2}$$

with a homogeneous function $G$ of $S_i$, satisfying the "transversality" condition such that the only solution for the simultaneous equations [5]

$$0 = \frac{\partial G}{\partial S_1} = \frac{\partial G}{\partial S_2} = \cdots = \frac{\partial G}{\partial S_n} \tag{5.3}$$

---

[6]As has been established in ref. [8], the $N = (2,0)$ global world-sheet supersymmetry is a necessary and sufficient condition for the $D = 4$, $N = 1$ space-time supersymmetry, which is the most interesting case phenomenologically. However, due to the complicatedness of these unidexterous models, we dealt only with the $N = (2,2)$ global supersymmetry in this paper.

[7]In this paper, we do not distinguish the symbols for component fields from those for superfields, as long as they are clear from the context.

is $\forall S_i = 0$. This guarantees that the smoothness of the hypersurface $G = 0$ in the space of the variables $S_i$. Now the bosonic potential (3.2) with the couplings to $N = (1,1)$ supergravity, with the $\theta$-term ignored, is

$$e^{-1}\mathcal{L}'_{\text{Pot}} = -|G(S_i)|^2 - |P|^2 \sum_i \left|\frac{\partial G}{\partial S_i}\right|^2$$
$$- \frac{\widetilde{e}^2}{2}\left(\sum_i |S_i|^2 - n|P|^2 - r\right)^2 - 2|\sigma|^2\left(\sum_i |S_i|^2 + n^2|P|^2\right) \quad (5.4)$$
$$- \frac{\widetilde{e}^2}{2(\sigma^1)^2}\left[P(G+G^*) - \sigma^2\left(\sum_i |S_i|^2 - n|P|^2 + r\right)\right]^2,$$

where the terms indicating the auxiliary field equations are skipped now. Except for the last line yielding the additional condition, all other terms in (5.4) are precisely the same as the global case [5].

We now perform an analysis similar to the global case [5], depending on the value of $r$. We also have to confirm the compatibility of the additional condition with the v.e.v.s in the global case [5]. This turns out to be rather simple, because, as was already mentioned, the second term in the last line in (5.4) is almost the same as the third term in (5.4) *except* the sign in the $r$-term, and therefore the vanishing of the first and the third terms in (5.4) will be consistent with the vanishing of its last term only when $\sigma^2 = 0$, as long as $r \neq 0$. This consideration gives the rough idea of the system, but there is subtlety about the singularity at $\sigma^1 \to 0$. We now check this more explicitly.

If $r \gg 0$, the vanishing of the third term in (5.4) tells that $\exists S_i \neq 0$. This implies that the second term in (5.4) vanishes, only when $P = 0$ due to the transversality condition. This with the third term of (5.4) fixes $\sum_i |S_i|^2 = r$. The first term in (5.4) can vanish, only when $G(S_i) = 0$. At this stage the first, second, third terms in (5.4) are vanishing with the v.e.v.s: $G(S_i) = 0$, $P = 0$, $\sum_i |S_i|^2 = r$. Now using these values in $\mathcal{L}'_{\text{Pot}}$ (5.4), we get

$$e^{-1}\mathcal{L}'_{\text{Pot}} = -r\left[(\sigma^1)^2 + (\sigma^2)^2\right] - 2\widetilde{e}^2 r^2\left(\frac{\sigma^2}{\sigma^1}\right)^2. \quad (5.5)$$

The last term is the effect of the extra $S$-auxiliary field. In order to avoid the subtlety around $\sigma^1 = 0$, we first minimize the last term with respect to $\sigma^2$, keeping $\sigma^1$ fixed. Eventually the last term vanishes at $\sigma^2 = 0$, and we are left only with the first term minimized at $\sigma^1 \to 0$. To put it differently, our potential (5.5) is rewritten as $V = 2r|\sigma|^2 + 2\widetilde{e}^2 r^2 \tan^2(\arg \sigma)$. so that the angle $\varphi \equiv \arg \sigma$ should be kept away from $\pi/2$, while approaching $|\sigma| \to 0$. The degree of the hypersurface $\{G = 0\} \subset \mathbb{CP}^{n-1}$ is the same as the degree $n$ of G. This implies that this hypersurface is equivalent to a CY manifold. Thus we see that the model on the curved world-sheet is also valid describing the target CY manifold.

We now see the case $r \ll 0$. In this case, the vanishing of the third term in (5.4) implies that $P \neq 0$, which combined with the second term in (5.4) means that $|\partial G/\partial S_i|^2 = 0$. Due to the transversality, this implies that $S_i = 0$, which now fixes $|P| = \sqrt{-r/n}$. The first term in (5.4) vanishes at $G(S_i) = 0$. Substituting these v.e.v.s in (5.4) results in a form similar to (5.5) with $r$ in its first term now replaced by $n|r|$. Therefore, we see that the potential is again minimized at $\sigma^2 = 0$ and $\sigma^1 \to 0$, in order to avoid the singularity. Interestingly, the vacuum structure of the model stays the same, such as the uniqueness of the vacuum up to gauge transformations, characterizing the LG model. We therefore see that this model is also valid as a LG model on the curved two-dimensional base manifold.

We see that at least these two cases of $r$, the minimization of the bosonic potential is realized at the zero value. Accordingly, we also see from (3.2) that all the auxiliary fields have zero v.e.v.s. The $N = (2,2)$ supersymmetry stays unbroken, as is expected also from the topological Witten index [5].

We finally mention the case of $N = (1,0)$ supergravity couplings. These couplings are much easier, because the auxiliary field $S$ is now absent, and there is no particular effect by the $N = (1,0)$ supergravity except for $\widetilde{\theta} = 0$ besides the usual gravitino and zweibein couplings compared with the global case [5].

## 6. Concluding Remarks

In this paper, we have examined the effect of $N = 1$ local supersymmetry on the global $N = (2,2)$ CY hypersurface sigma-models, which are important for explicit model building for superstring theories as well as LG theory. It turned out that the only effect by the $N = (1,1)$ supergravity is through the auxiliary field $S$. In the case of $N = (1,1)$ supergravity couplings, there is no essentially new effect on the geometry of the hypersurface sigma-model by the $D = 2$ geometry except for the extra condition together with the condition $\widetilde{\theta} = 0$. The case of $N = (1,0)$ supergravity is even simpler, because the auxiliary field $S$ is also absent.

As mentioned in the introduction, making only the $N = (1,1)$ and $N = (1,0)$ supersymmetries local is motivated respectively by the $N = 2$ superstring theory [7] and $N = 1$ superstring or heterotic string theory [2][8]. To our knowledge, our result is the first explicit one for such couplings of supergravity to the $N = (2,2)$ global supersymmetric system, including the minimal $U(1)$ couplings of vector multiplets. It also supports the validity of applying such sigma-models as superstring theory to phenomenological model building with practical features of CY manifolds.

We have seen that the $N=1$ local supergravity couplings [9] to sigma-models in $D=2$ especially with the $U(1)$ minimal couplings has some features different from the analogous $D=4$ case [11], such as the absence of the Brans-Dicke type term [9]. This resulted in the independence of the curvature of the world-sheet, which might have spoiled the geometrically nice features of the global case in the sigma-model interpolating between the CY and LG models we have analyzed. The basic geometrical structure persists at any genus of the string world-sheet, and there is no particular effect by its "curvedness" on the CY hypersurface in the target space.

In this paper we have dealt only with the CY hypersurface sigma-model sector in the total superstring model, e.g., in $N=1$ heterotic string theory, there are other sectors such as for the $D=10$ purely supergravity sector, or the purely "matter" sector corresponding to the $D=10$ supersymmetric Yang-Mills multiplet. This is because the general couplings in these sectors have been already established in the past [13][14]. The peculiar effect arises, when there are some $U(1)$-gauging for the $N=(2,2)$ hypersurface sigma-models, especially in the presence of $N=(1,1)$ supergravity, which were not successful in the past [13].

Once we have established the $U(1)$-gauging, it is straightforward to generalize it to the non-Abelian case. This can be easily done based on the similar $D=4$, $N=1$ case [11] as a guiding principle. We give here only the results by stating how the generalization goes with relevant terms. All the terms in (2.11) are to be replaced by the non-Abelian ones, such as the kinetic terms $-(1/4)(V_{\mu\nu}^I)^2$ with the indices $I, J, \cdots$ for the adjoint representation of the non-Abelian gauge group. All other bilinear terms there are also replaced by the contraction with respect to these $I, J, \cdots$ indices. Now the chiral multiplet lagrangian (2.8) is accordingly generalized as usual, such as the $\phi$-kinetic term now contains the non-Abelian covariant derivative, etc. The most important change occurs in the $Q$-explicit terms of the last three lines in (2.8) generalized to the expression:

$$\mathcal{L}_{\text{CM, Q}} = \bigg[ -\sqrt{2}Q(\lambda^I T^I \psi)^i \phi_i^* - \sqrt{2}iQ(\overline{\psi}_\mu \gamma^\mu T^I \psi)^i \phi_i^* \sigma^I$$
$$- \frac{i}{\sqrt{2}} Q\overline{\psi}_\mu \gamma^\mu \lambda^I (\phi T^I \phi^*) - \frac{1}{\sqrt{2}} \overline{\psi}_\mu \gamma^{\mu\nu} \psi_\nu (\phi T^I \phi^*) \sigma^I \bigg] + \text{h.c.} \tag{6.1}$$
$$- Q\sigma^{1I}(\psi T^I \gamma_5 \overline{\psi}) - iQ\sigma^{2I}(\psi T^I \overline{\psi}) - iQ(\phi T^I \phi^*) D^I + 2Q\sigma^I \sigma^{*J}\left(\phi\{T^I, T^J\}\phi^*\right) \ .$$

The indices $i, j, \cdots$ on the chiral multiplet are now for an arbitrary representation, while $T^I$ are the anti-hermitian generators of the gauge group: $\left[(T^I)_i{}^j\right]^* = -(T^I)_j{}^i$, acting as

$$(\phi T^I \phi^*) \equiv \phi^i (T^I)_i{}^j \phi_j^* \ . \tag{6.2}$$

The $Q$ is now the gauge coupling, while the *bars* on the Weyl spinors such as $\overline{\psi}^i$ are as mentioned before, and the Weyl spinor $\lambda$ is defined by $\lambda \equiv (\lambda^1 + i\lambda^2)/\sqrt{2}$ like $\psi$.

In this paper we have dealt only with Poincaré supergravity couplings to the hypersurface sigma-models [5]. This is because the latter models originally lack the superconformal invariance, *except for* the infrared fixed points where the superconformal invariance is restored. From this viewpoint it is also interesting to see the quantum effect of the supergravity couplings.

There are many applications of our result, owing to its general structure with the $U(1)$-gauge couplings, and its importance and relevance to the realistic model building based on CY space with local $N = 1$ supersymmetry on the world-sheet in superstring theory. It seems worthwhile to study sigma-models for CY hypersurfaces in Grassmannian [6] or gauged LG models by the help of the non-Abelian couplings with local supersymmetry we developed. Additionally we can seek possible mechanisms of non-perturbative supersymmetry breakings as in refs. [12][17][18][19].

We are indebted to S.J. Gates, Jr., T. Hübsch and E. Witten for valuable suggestions and for reading the manuscript.